# Significance of Side Information in the Graph Matching Problem

Kushagra Singhal[1] Daniel Cullina Negar Kiyavash

*University of Illinois, Urbana IL 61801, USA*
ksingha2@illinois.edu

Abstract. Percolation based graph matching algorithms rely on the availability of seed vertex pairs as side information to e ciently match users across networks. Although such algorithms work well in practice, there are other types of side information available which are potentially useful to an attacker. In this paper, we consider the problem of matching two correlated graphs when an attacker has access to side information, either in the form of community labels or an imperfect initial matching. In the former case, we propose a naive graph matching algorithm by introducing the community degree vectors which harness the information from community labels in an e cient manner. Furthermore, we analyze a variant of the basic percolation algorithm proposed in [44] for graphs with community structure. In the latter case, we propose a novel percolation algorithm with two thresholds which uses an imperfect matching as input to match correlated graphs.

We evaluate the proposed algorithms on synthetic as well as real world datasets using various experiments. The experimental results demonstrate the importance of communities as side information especially when the number of seeds is small and the networks are weakly correlated.

## 1 Introduction

Graph matching algorithms have become a focus of research among various communities recently [20, 21, 22, 31, 45, 25]. Given two graphs, $G_1$ and $G_2$, with correlated vertex and edge sets, a graph matching algorithm tries to nd a mapping between the vertex sets of $G_1$ and $G_2$. Graph matching algorithms nd relevance in different domains. An example, which is the focus of this work, is the study of privacy risks involved in sharing data of social network users [11, 2, 41]. Each user in a social network can be represented as a vertex, and an interaction (friendship, message etc.) between two users is indicated by an edge. The shared data helps data mining researchers in understanding network related properties [39, 40, 37, 3, 32, 42, 26] as well as organizations to generate revenues using product advertisements and recommendations [15, 46, 28]. Although of great utility, this data has the potential to leak user privacy [24, 29] and this privacy-utility trade-off has been noted in the literature [7, 27]. Another example is the alignment of protein-protein interaction (PPI) networks, which is an important step in understanding the biological processes involved in cell interactions [38]. In this case, vertices of a graph represent proteins and a direct physical interaction between two proteins is represented by an edge.

The two networks, $G_1$ and $G_2$, may represent two different social networks, for example Facebook and Google+, or may even be the snapshots of the same network at two different times. The target network, $G_2$, often contains private information about network users, while real identities may be known in the auxiliary network, $G_1$ [19]. Thus, a correct mapping between the vertex sets of the two graphs results in associating private information in $G_2$ to real world individuals, which is a breach of user privacy. Such an identication of individuals in a network is termed network deanonymization.

Previous studies, for the most part, focused on availability of the, so called, seed vertex pairs to match correlated graphs. For example, Narayanan and Shmatikov [31] successfully matched a fraction of users in two real world networks using a heuristic algorithm. Nilizadeh et al. [34] incorporated the commu-nity level information to improve upon this work. Although Nilizadeh et al. use the community structure of the networks, our work diers from their's in several ways. We analyze the importance of community structure in the graph matching problem when the seeds are not available to an attacker. For this purpose, we propose an algorithm which uses the information from the community structure by utilizing the community level degrees of the vertices, rather than the global degrees. Also, the structure of the percolation algorithms, proposed in this paper, is different than the algorithm in [34]. We would like to emphasize that the results in this paper, in fact, complement the results in [34] by providing a theoretical understanding of the importance of the community structure in graph matching problem. Wondracek et al. [43], solving a slightly different problem, utilized the group membership information to identify users in a network. Some studies also approached this problem from a theoretical viewpoint. For example, Yartseva and Grossglauser [44] proposed and analyzed a graph matching algo-rithm based on bootstrap percolation. Chiasserini et al. [8] studied the Effect of clustering on network deanonymization and showed empirically that clustering can potentially decrease the initial seed set size required to percolate in random geometric graphs.

**Contributions.** All the aforementioned work on network deanonymization focused only on one type of side information, i.e., seeds. This motivated us to ask the following question: If an attacker has access to other side information, including seeds, how could it be used eciently in network deanonymization? The main aim of this paper is to investigate the role played by the side information in the graph matching problem. We will consider three types of side information in this paper: seeds, community structure, and an imperfect matching of the graphs $G_1$ and $G_2$. We would like to emphasize that the source of these side information is not the concern of this work, though it would be an interesting research direction.

To analyze the problem theoretically, we consider two correlated graphs, $G_1$ and $G_2$, generated using the Stochastic Block Model (SBM) [17]. The contribution of this work is twofold. First, we propose and analyze a naive algorithm which uses community degree vectors of the network users to deanonymize them. Assuming that the number of communities grows slowly with network size, we derive conditions on model parameters such that it is possible to match the graphs perfectly using this algorithm. We also propose and analyze a variant of simple percolation algorithm [44] for graphs with community structure. We show that the matching threshold [44] r = 2 is su cient to match almost all users correctly when the number of communities grows faster that the square of the logarithm of the number of vertices. Second, we propose a two-threshold percolation algorithm which uses an imperfect matching as an input. The imperfect matching is used to identify pairs of users which have high probability of being correct pairs. These special pairs are assigned a lower threshold for the percolation step, while the rest of the pairs are assigned a higher threshold. The use of different thresholds for different kinds of pairs is useful in two ways. When the correct pairs are more probable to get a lower threshold, the number of seeds required to percolate decreases. Also, as the percolation algorithms tend to make more errors in the beginning, mapping the correct pairs early reduces the overall error of the algorithm. Assuming that the algorithm percolates, we derive conditions under which the algorithm does not make an error with high probability.

We conduct experiments on synthetic as well as real world networks by varying the number of seeds, levels of correlation, and fraction of community labels known to the attacker. Based on our results, we nd that availability of side information in the form of community labels is very useful when (i) the number of seeds is small, and (ii) the correlation between the datasets is weak. We also nd that the availability of community labels for a fraction of users is enough for the proposed percolation graph matching algorithm to perform better than other similar algorithms.

In the end, we discuss some implications of our work and experimental results for network privacy. We also discuss possible future directions and questions that need to be answered in order to design better techniques to mitigate the Effects of side information in network deanonymization.

The rest of the paper is organized as follows. We begin by reviewing the relevant literature in Section 2. The system and the attack models are given in Section 3. Section 4 describes the proposed algorithms and their theoretical analysis. The experimental results are discussed in Section 5. We conclude with some remarks and future directions in Section 6.

## 2 Literature Review

The network deanonymization problem has received attention in the recent years. The problem has mostly been studied from a practical viewpoint but some the-oretical aspects have also been studied. In this section, we discuss some of the important literature associated with the problem.

Narayanan and Shmatikov [31] were among the rst researchers to study the network deanonymization problem. They proposed a heuristic algorithm to deanonymize users of a network when an adversary has access to a correlated auxiliary network. The algorithm rst identi ed some seed vertex pairs and then used this information to propagate and map the remaining users. They were successful in mapping a fraction of the users of Twitter and Flickr networks with small error rate. Nilizadeh et al. [34] extended this method and incorporated the community structure of the networks. Utilizing the community structure reduced the size of candidate user pairs and helped in identifying more seed ver-tices, thereby reducing the error rate. They were able to map correlated large Twitter networks with as few as 16 seeds. They also noted that their method only works if the correlation between the networks is high. Wondracek et al. [43] consider the problem of deanonymizing users when their communities' member-ships are known in a network. They also created community ngerprints of users by stealing their browser history. They showed, by comparing ngerprints and memberships, that some of the users could be identi ed uniquely while the can-didate set is greatly reduced for others. Although these works give insight into the problem of deanonymization, they lack a theoretical framework and seem to be dataset speci c.

Some e cient deanonymization algorithms have been proposed recently. Ji et al. [20] analyzed the graph matching problem for the con guration model. They proposed an optimization based deanonymization algorithm to match the two graphs, but did not provide theoretical guarantees for the algorithm. Yartseva and Grossglauser [44] proposed and analyzed a percolation algorithm for match-ing correlated Erd•os-Renyi graphs. The algorithm begins by considering a set of seed pairs which spread their marks to other pairs of vertices. As soon as a par-ticular pair of vertices reaches a xed threshold, r, it is considered as matched. They established a phase transition in the size of initial seed set for matching the graphs almost completely. There are two main drawbacks to their method. First, the theoretical guarantee is only proved for r 4, which makes the algorithm impractical as it requires a huge number of seeds in that case. Second, the threshold r is same for all vertex pairs. In [22], Kazemi et al. proposed a variant of percolation algorithm which requires fewer initial seeds but the error increases slightly compared to the former algorithm. They also showed that r = 2 suffices to match most of the vertex pairs with only o(n) pairs matched incorrectly. To overcome the shortcomings of [44, 22], we propose a two-threshold percolation graph matching algorithm. As will be seen later, the proposed algorithm not only percolates with fewer seeds but also achieves lower error rates compared to the algorithms in [44, 22].

Recently, some information theoretic results were derived for matching corre-lated graphs drawn from random models. Pedarsani and Grossglauser [36] ana-lyzed the graph matching problem theoretically for Erd•os-Renyi random graphs and derived an achievability result for matching two correlated graphs perfectly.

They established that the average degree has to grow slightly faster than the logarithm of the size of the graphs in order to achieve perfect deanonymization. Cullina and Kiyavash [9] improved upon their result for achievability and also derived an almost tight converse for the problem. In [10], Cullina et al. proved similar achievability and converse results for the graphs drawn from the Stochastic Block Model. They also derived conditions on model parameters which enable perfect community recovery while preventing perfect de-anonymization simultaneously. Onaran et al. [35] also derived an achievability result for this problem when the network is divided into two possibly unequally sized communities. Ji et al. [19] considered the exact as well as partial deanonymization problem when seed vertex pairs are available to the attacker. They derived achievability bounds for Erd•os-Renyi random graphs and extended the results to more general graphs. Although these information theoretic results provide good insights into the problem, they do not provide e cient algorithms to deanonymize networks.

Fabiana et al. [13] proposed a deanonymization algorithm for matching graphs with power law degree distribution. They used degree distribution to divide the user pairs into categories. Based on their degrees, the matching thresh-olds are assigned to these pairs. The algorithm rst uses higher threshold to match moderately higher degree users, then progressively reduces the threshold to match low degree users, and in the end matches users with very high degrees. Although they use different thresholds for different pairs, the minimum thresh-old used is $r_{min} = 3$, and as seen in their experiments, the algorithm requires many seeds to percolate when the scale-free graphs are moderately correlated. In contrast, our algorithm requires far fewer seeds to percolate even when the correlation between the datasets is weak.

## 3 The Graph Matching Problem

This section introduces, in detail, the graph matching problem that the paper aims to solve. First, representation of social networks as graphs is discussed followed by the description of the Stochastic Block Model (SBM) for random graphs. Next, the process of generating correlated graphs is explained along with the attack model.

### 3.1 Social Networks as Graphs

As noted in previous literature [31, 34, 36, 23], a social network can be modeled as a graph (directed or undirected). In this paper, the vertex and the edge sets of a graph G are denoted by V(G) and E(G) respectively. The users of a social network are represented as vertices in a graph whereas the existence of a relation (e.g. friendship link, messages etc.) between two users is modeled by an edge between them. For the purpose of this paper, social networks are modeled as undirected graphs (directed graphs contain more information and are easier to deanonymize [19]).

## 3.2 Stochastic Block Model

Communities are an integral part of any social network [14, 30]. Community detection in networks [16, 1, 14] is one of the most widely studied problem in social network analysis. The Stochastic Block Model (SBM) is a random graph model used to generate graphs with community structure [17].

Consider an n vertex graph, G, with vertex set V (G) and edge set E(G). Also, let there be a community label assignment function *C: V (G)→ {1, 2,,,,,, K}* $\triangleq$ *[K]*, where K is the number of communities in G. This function associates a community label with every vertex in G and hence partitions V (G) into disjoint sets called communities. Let $C_k \triangleq$ *{v ε V (G): C(v) = k}* denote the vertex set of the k[th] community. We assume that communities are equally sized[1], i.e., $|C_i| = |C_j|$ for *i, j* ε *[K]*. Let P(X) denote the probability of an event X. The edge set, E(G), is generated as follows. For vertices *u, v ε V (G),* we have

$$P(uv \in E(G)) = \begin{cases} p & \text{if } \mathcal{C}(u) = \mathcal{C}(v) \\ q & \text{if } \mathcal{C}(u) \neq \mathcal{C}(v) \end{cases}$$

where p, q ε (0, 1) and are allowed to depend on n. Note that, p and q denote the probability that an edge exists between vertices belonging to the same com-munity and different communities, respectively. In this model, the existence of an edge is independent of the existence of any other edge. We assume that p q for the rest of this work as the communities are assortative in real networks. The graph distribution associated with this generative method is denoted by SBM (n, p, q, K). When p = q or K = 1, this model degenerates to the Erd•os-Renyi random graph model [12].

Although the Stochastic Block Model is not completely representative of the community structure found in real world networks, it is used often in the theoretical analysis for problems involving communities in graphs. The main reasons for this choice are as follows. First, it captures the assortativity property exhibited by the real networks. Second, the theoretical analysis of the problem becomes tractable, and also results in providing an intuitive explanation of the problem for real networks.

## 3.3 Problem Formulation

**Generating Correlated** Graphs We use the following model to generate correlated graphs. Consider an underlying graph G with vertex set V (G) = {1, 2,,,,,,,n} $\triangleq$ [n] and G~SBM (n, p, q, K). First, generate graph G*1* with V (G*1*) = V (G). The edge set, E(G*1*), is then generated as follows. For every edge e 2 E(G), P (e ε E(G*1*)) = s, where s ε (0,1) is called the sampling parameter, and P (e ε E(G*1*)) = 0 if e $\notin$ E(G). The sampling of every edge is independent of other edges. The graph $G_2$ is generated independently and identically. Figure 1 depicts a toy example of this generating process. Note that this is the same model which was used in previous studies [19, 21, 22]. We further assume that the vertices of $G_2$ are anonymized using naive anonymization in which any personal identi ers are replaced by random identi ers. Although it is a simple anonymization technique, naive anonymization is still the most widely used method to anonymize datasets before publishing [31, 41]. Hence, we index the vertices of $G_2$ by $\{1', 2', ....., n'\} \triangleq [n']$. Without loss of generality, we assume that $i \in V (G_1)$ and $i' \in V (G_2)$ correspond to the same vertex in G. This notation proves bene cial for the analysis later. It is useful to note that G1, G2 ~ SBM (n, ps, qs, K). Let the vertex set of the Kth community in $G_1$ and $G_2$ be denoted by $C_k^1$ and $C_k^2$ respectively. More formally, for $z \in \{1,2\}$, $C_k^z = \{v \in V(G_2)\ ,\ C(v) = k\}$. To ensure connectivity [12] in the graphs $G_1$ and $G_2$ we consider graphs with logarithmic average degree, i.e., $p = a\frac{\log n}{n}$ and $q = b\frac{\log n}{n}$ , where a; b are constants which are called model parameters. This choice of average degree also captures the sparse nature of connectivity found in the real world networks. Also, we assume that the number of communities is $K = n^\alpha$ (unless otherwise stated), $\alpha \in (0, 1)$, which means that the number of communities grows with number of vertices sublinearly.

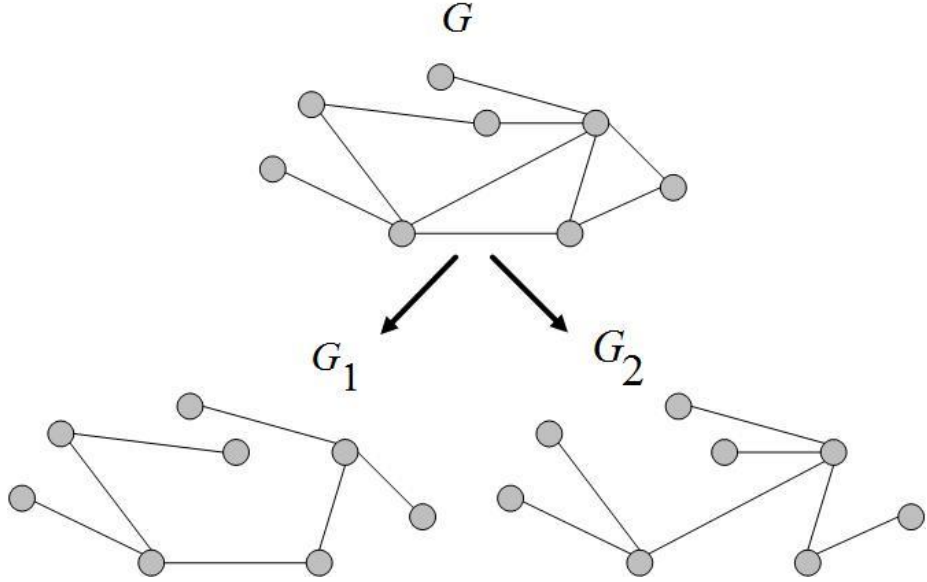

Fig. 1: A toy example for generating two correlated graphs.

**Modeling the Attacker** An attacker is any entity that aims to match the vertices of $G_1$ to those of $G_2$, which in turn breaches the privacy of the users in a social network. More formally, the goal of an attacker is to construct a bijective mapping $\pi$: V ($G_1$) → V ($G_2$) such that $\pi = \pi_0$ where $\pi_0(i) = i'$ for all i ∈ V (G$_1$). We call $\pi_0$ the ground truth mapping. For the graph G$_1$, let $\pi\big(E(G_1)\big) \triangleq \{\pi(u)\,\pi(v)\,,\,u,\,v \in V(G_1),\,uv \in E(G_1)\}$. Also, let $\pi^{-1}$, V (G$_2$) → V (G$_1$) denote the inverse of mapping $\pi$.

   We consider an attacker with one or more of the following three possible side informations.

-Set of seed vertices, S$_0$, defined as

$$S_0 = \{(i;\,i^{'}) : i \in V(G_1),\, i^{'} \in V(G_2)\} \tag{1}$$

where |S$_0$| = Φ. This set provides an attacker with vertex pairs which are already matched in the two graphs.

- Community labels of all vertices in both graphs. For a vertex i ∈ V(G$_1$), this information restricts the set of candidate vertices to $C^2_{C(i)}$. In Section 5, we show that even partial information about the community labels is helpful in graph matching.

- An imperfect matching of the graphs, in which, each vertex i ∈ V(G$_1$) is assigned a vertex j$^{'}$ ∈ V(G$_2$). The pair (i, j$^{'}$) is said to be matched by this imperfect matching. The term imperfect is used to emphasize the fact that the attacker does not know which of the pairs are correctly matched.

*Remark 1.* We will only consider mappings which preserve the community structure when we assume that an attacker knows the community labels of the users. The reason is that any rational attacker would use this information to limit the search space for the matching.

## 4 Graph Matching Algorithms with Side Information

This section discusses the main contributions of the paper. We discuss three graph matching algorithms. The first, called naive algorithm, leverages the community label information and does not require seeds to match users across two correlated graphs. This is different than most of the algorithms proposed in the literature, which depend on seeds to match users further. We derive an achievability bound for this algorithm, and thus identify the parameter space where the algorithm matches the graphs perfectly with high probability. The second al-gorithm is an extension of the basic percolation algorithm in [44] to graphs with community structure. We show that when the community labels are known, the graph matching task becomes much easier and the error-free matching threshold can be as small as r = 2 when the number of communities is K = ώ(log² n). The third algorithm assumes that an imperfect matching of the users is given as input and uses this information along with a modi ed graph matching per-colation algorithm with two thresholds. We derive conditions under which the imperfect information helps the percolation algorithm to match users in the two graphs without making errors.

### 4.1 The Naive Algorithm

Consider two correlated SBM graphs, G$_1$ and G$_2$, generated as in Section 3.3. For some $k_0 \epsilon$ [K], consider a vertex i $\epsilon$ $C_{k0}^1$. For each k $\epsilon$ [K], let d$_i$(k) = |{j $\epsilon$ C$_k^1$ : ij $\epsilon$ E(G$_1$)}| be the size of the neighborhood of i in community k. Similarly, for a vertex j $\epsilon$ $C_{k0}^2$ define d$_j$(k). Then the community degree vector of i, denoted by d$_i$, is defined as

$$d_i = [d_i(1),\ldots\ldots,d_i(k_0 \text{-}1),\, d_i(k_0 + 1),\ldots\ldots,d_i(K)]^T \qquad (2)$$

Similarly, let d$_{j0}$ denote the community degree vector of j$^0$. The community degree vectors represent the pattern in which a user connects to various communities in the network. If the number of communities is big enough, there is a nontrivial amount of information in these vectors which can be utilized to deanonymize the network. In particular, when the graphs are sparse, users are connected to only a few communities, which results in enough information to deanonymize the users in the network. Intuitively, we expect d$_i$ to be closer to d$_i$ than d$_j$ for j$^{'} \neq$ i$^{'}$. We de ne the distance between the pair (i, j$^{'}$) as

$$\Delta_{ij'} = \sum_{k=1, k \neq k_0}^{K} \Delta_{ij'}(k) \qquad (3)$$

$$\Delta_{ij'}(k) = \mathbf{1}(d_i(k) \neq d_{j'}(k)). \qquad (4)$$

This distance measures the dissimilarity of the community degree vectors of the vertices i and j$^{'}$. If both i and j$^{'}$ have the same number of neighbors in a particular community, the contribution of this community to the distance in (3) is 0. In all other cases, the contribution is 1.

We now propose a naive algorithm to match the graphs G$_1$ and G$_2$. This algorithm works as follows. For every vertex i $\epsilon$ V (G$_1$), the algorithm returns j'$_i$ $\epsilon$ V (G$_2$) such that i and j'$_i$ have the same community label and j'$_i$ is closest to i in terms of the distance defined in (3), that is, $j'_i = \mathrm{argmin}_{j' \in C_{c(i)}^2} \Delta_{ij'}$. The pseudo code is described in Algorithm 1. This algorithm is conceptually simple and computationally efficient as most of the steps in the algorithm could be performed in parallel.

The rest of this section is aimed at deriving conditions under which the naive algorithm matches the graphs G$_1$ and G$_2$ perfectly with high probability. An important observation is that the individual terms in (3) are independent because they correspond to edges to different communities. Also, they are identically distributed. We are interested in the conditions on the model parameters such that P ( $\Delta_{ii} \geq \Delta_{ij'}$) $\rightarrow$ 0 as n $\rightarrow$ ∞ for all i and j$^{'}$. This means that all correct pairs of vertices are closer to each other, in terms of distance in (3), compared to incorrect pairs, asymptotically with high probability.

First, we bound the probability that an arbitrary incorrect pair of vertices are closer to each other than the correct pair in terms of distance in (3).

**Lemma 1** *For any* $k_0 \in [K]$, $i \in C_{k_0}^1$, $j' \in C_{k_0}^2$, $j' \neq i'$, *and* $s > 0$,

---
**Algorithm 1** Naive Algorithm
---
**Require:** $G_1, G_2, \mathcal{C}$

    Compute the community degree vectors, $\mathbf{d}_i$ and $\mathbf{d}_{j'}$, for all vertices $i \in V(G_1)$
    and $j' \in V(G_2)$

    **for** all vertex pairs $(i, j') \in V(G_1) \times V(G_2)$ such that $\mathcal{C}(i) = \mathcal{C}(j')$ **do**

        Calculate $d_{ij'}$ using Equation (3)

    **end for**

    **for** $i \in V(G_1)$ **do**

        $j'_i = \mathrm{argmin}_{j' \in C^2_{\mathcal{C}(i)}} d_{ij'}$

        Return $(i, j'_i)$ as a matched pair

    **end for**
---

$$P[\Delta_{ii'} \geq \Delta_{ij'}] \leq n^{-2bs(1-\sqrt{1-s^2})+o(1)} \to 0 \ \ as \ \ n \to \infty$$

*Proof.* We need to analyze $P(\Delta_{ii'} - \Delta_{ij'} \geq 0)$.

    Let $\widetilde{\Delta}(k) = \mathbf{1}(d_i(k) \neq d_{i'}(k)) - \mathbf{1}(d_i(k) \neq d_{j'}(k))$. We have

$$P[\Delta_{ii'} - \Delta_{ij'} \geq 0]$$

$$= P\left[\sum_{k=1, k \neq k_0}^{K} \mathbf{1}(d_i(k) \neq d_{i'}(k)) - \mathbf{1}(d_i(k) \neq d_{j'}(k)) \geq 0\right]$$

$$= P\left[\sum_{k=1, k \neq k_0}^{K} \widetilde{\Delta}(k) \geq 0\right] \tag{5}$$

Note that $\widetilde{\Delta}(k)$'s are independent and identically distributed. Let $p_i = P(\widetilde{\Delta}(k) = i)$, $i \in \{-1, 0, 1\}$. Let $E(Z)$ denote the expected value of the random variable $Z$. For any $x > 0$,

$$P(\sum_{k=1, k \neq k_0}^{K} \widetilde{\Delta}(k) \geq 0) = P(e^{x \sum_{k=1, k \neq k_0}^{K} d(k)} \geq 1)$$

$$\overset{(a)}{\leq} \mathbb{E}(e^{x \sum_{k=1, k \neq k_0}^{K} d(k)}) = [\mathbb{E}(e^{xd(k)})]^{K-1}$$

$$= (p_0 + p_1 e^x + p_{-1} e^{-x})^{K-1}$$

$$\overset{(b)}{\leq} \exp((K-1)(p_0 + p_1 e^x + p_{-1} e^{-x} - 1))$$

$$\overset{(c)}{\leq} \exp(-(K-1)(\sqrt{p_1} - \sqrt{p_{-1}})^2) \tag{6}$$

where (a) follows from the Markov's inequality, (b) holds because $y \ e^y \ \mathbf{1}$ for any y, and (c) follows after substituting the optimal value of x.

    Now we need to calculate $p_1$ and $p_{-1}$. Also, let $D_i(k)$ denote the number of neighbors of vertex $i$ in community $k$ in the underlying graph $G$ from which $G_1$ and $G_2$ were edge sampled. Then,

$$p_1 = P(\widetilde{\Delta}(k) = 1)$$

$$= P(\Delta_{ii'}(k) = 1, \Delta_{ij'}(k) = 0)$$

$$\overset{(a)}{=} \sum_{m_i=1}^{n/K} \sum_{m_j=0}^{n/K} P(D_i(k) = m_i, D_j(k) = m_j) \times$$

$$P(\Delta_{ii'}(k) = 1, \Delta_{ij'}(k) = 0 | D_i(k) = m_i, D_j(k) = m_j)$$

$$\overset{(b)}{=} P(\Delta_{ii'}(k) = 1, \Delta_{ij'}(k) = 0 | D_i(k) = 1, D_j(k) = 0) \times$$

$$P(D_i(k) = 1, D_j(k) = 0) + o\left(\frac{nq}{K}\right)$$

where (a) follows from the law of total probability. To prove (b), we use the following chain of equations. First, using $\binom{n/K}{m_i} < \left(\frac{n}{k}\right)^{m_i}$, we have

$$\sum_{m_i=1}^{n/K} \sum_{m_j=1}^{n/K} P(D_i(k) = m_i, D_j(k) = m_j)$$

$$\leq \sum_{m_i=1}^{n/K} \sum_{m_j=1}^{n/K} \left(\frac{nq}{K}\right)^{m_i+m_j}$$

$$= \left(\frac{\frac{nq}{K}\left(1 - \left(\frac{nq}{K}\right)^{\frac{n}{K}}\right)}{1 - \frac{nq}{K}}\right)^2$$

$$\leq \left(\frac{nq}{K}\right)^2 (1 + o(1))$$

$$= o\left(\frac{nq}{K}\right)$$

Now, (b) follows because any conditional terms can be bounded above by 1. So,

$$p_1$$
$$= P(d_i(k) = d_{j'}(k) = 0, d_{i'}(k) = 1 | D_i(k) = 1, D_j(k) = 0)$$

$$\times P(D_i(k) = 1, D_j(k) = 0) + o\left(\frac{nq}{K}\right)$$

$$= s(1-s)\frac{nq}{K}(1-q)^{\frac{2n}{K}-1} + o\left(\frac{nq}{K}\right)$$

$$= s(1-s)\frac{nq}{K}\left(1 - o\left(\frac{nq}{K}\right)\right) + o\left(\frac{nq}{K}\right)$$

$$= s(1-s)\frac{nq}{K} + o\left(\frac{nq}{K}\right)$$

$$(7)$$

Similarly,

$$p_{-1}$$
$$= P(\widetilde{\Delta}(k) = -1)$$
$$= P(\Delta_{ii'}(k) = 0, \Delta_{ij'}(k) = 1)$$
$$= P(\Delta_{ii'}(k) = 0, \Delta_{ij'}(k) = 1 | D_i(k) = 0, D_j(k) = 1)$$
$$\quad \times P(D_i(k) = 0, D_j(k) = 1) + o\left(\frac{nq}{K}\right)$$
$$\quad + P(\Delta_{ii'}(k) = 0, \Delta_{ij'}(k) = 1 | D_i(k) = 1, D_j(k) = 0)$$
$$\quad \times P(D_i(k) = 1, D_j(k) = 0) + o\left(\frac{nq}{K}\right)$$
$$= s\frac{nq}{K}(1-q)^{\frac{2n}{K}-1} + s^2 \frac{nq}{K}(1-q)^{\frac{2n}{K}-1} + o\left(\frac{nq}{K}\right)$$
$$= s(1+s)\frac{nq}{K}\left(1 - o\left(\frac{nq}{K}\right)\right) + o\left(\frac{nq}{K}\right)$$
$$= s(1+s)\frac{nq}{K} + o\left(\frac{nq}{K}\right)$$

$$(8)$$

So we have

$$P\left[\sum_{k=1}^{K-1} \widetilde{\Delta}(k) \geq 0\right]$$
$$\leq \exp(-(K-1)(\sqrt{p_1} - \sqrt{p_{-1}})^2)$$
$$= \exp(-(K-1)\frac{2snq}{K}(1 - \sqrt{1-s^2+o(1)}))$$
$$= \exp(-2nqs(1 - \sqrt{1-s^2+o(1)}))$$
$$= \exp(-2bs(1 - \sqrt{1-s^2+o(1)})\log n)$$
$$\leq n^{-2bs(1-\sqrt{1-s^2})+o(1)}$$

$$(9)$$

The following theorem provides sufficient conditions to match $G_1$ and $G_2$ perfectly using only the community degree vectors.

**Theorem 1.** If $b > \frac{2-\alpha}{2s\left(1-\sqrt{1-s^2}\right)}$, and s>0, then the naïve algorithm matches the graphs $G_1$ and $G_2$ perfectly with high probability.

*Proof.* Let $P_e$ denote the probability that at least one vertex of $G_1$ is matched incorrectly. Note that, for every vertex i $\epsilon$ V ($G_1$), there are exactly $n^1$ candidates j'. Using Lemma 1 and union bound over i; j', we have

$$P_e \leq \bigcup_{i,j'} P(\Delta_{ii'} \geq \Delta_{ij'})$$
$$\leq n^{2-\alpha-2bs(1-\sqrt{1-s^2})+o(1)}$$
$$\to 0 \text{ if } b > \frac{2-\alpha}{2s(1-\sqrt{1-s^2})}$$

This result implies that when the inter-community edge probability is high enough (recall that $q = b \frac{\log n}{n}$), the community degree vectors contain enough information to deanonymize all the users in the network.

Note that the right-hand side of the condition in Theorem 1 is a decreasing function of s. When the value of s is large enough, the naive algorithm can deanonymize even very sparse graphs. On the other hand, when s is small (say s < 0.75), it requires denser graphs (large value of b) to deanonymize all the users. Even if the algorithm may not deanonymize all the users, a small fraction of users may still be deanonymized correctly by the naive algorithm. In such situations, the algorithm can be used as a preprocessing step to other algorithms such as the one described later in Section 4.2.

## 4.2 Percolation Algorithms

Percolation based graph matching algorithms have been recently proposed in literature [44, 22]. Such algorithms are computationally very e cient while providing good deanonymization results. In this section, we describe variants of the basic percolation algorithm in [44] for matching correlated graphs using side in-formation. We rst review the basic percolation algorithm introduced in [44], followed by the variant when an attacker has access to the community label in-formation. We analyze this algorithm and show that, when $K = \dot{\omega}(\log^2 n)$, the matching threshold r = 2 suffices to match almost all the vertices in the graphs correctly. We also propose a two-threshold percolation algorithm which uses an imperfect matching as input and utilizes this information to match further vertices efficiently.

**Basic Percolation Algorithm** The basic percolation algorithm [44] begins by considering a seed set $S_0$ defined in Equation (1). At every time step, a seed pair $(u; u')\in V(G_1) \times V(G_2)$ is selected from this set. For all potential vertex pairs $(i; j)\in V(G_1) \times V(G_2)$ such that $(i; u)\in E(G_1)$ and $(j; u)\in E(G_2)$, the score of the pair $(i; j^0)$ at time t, denoted by $M_{ij^0}(t)$, is increased by one. When the score of a pair reaches a threshold r, it is added to the seed set and all other pairs involving i and $j'$ are removed from further considerations. If the size of the initial seed set is large enough, then the algorithm percolates to n -- o(n) correct vertex pairs if r ≥ 4. For more details see [44].

**Percolation with Community Label Information** Suppose that an attacker knows the community labels of the users in graphs $G_1$ and $G_2$. We show that when the number of communities, K, is large enough, the basic percolation algorithm percolates to $\frac{n}{K} - o(\frac{n}{K})$ correct pairs in each community with high probability. We establish this by using a tighter bounding technique than that used in [44].

We consider a variant of the basic percolation algorithm, such that, instead of picking a seed at every time step, the algorithm picks one seed from each community at every time step. Thus, our algorithm picks K seeds at a time step. Our aim is to show that this algorithm does not match wrong pair of vertices with high probability even when r = 2. Hence, we will establish that the community label information helps in relaxing the threshold from r = 4 to r = 2 and thus the algorithm can percolate with small number of seeds. We also provide an upper bound on the critical number of seeds needed by our algorithm to percolate.

**Theorem 2**. Let the percolation threshold be r = 2. When the community labels of the users in $G_1$ and $G_2$ are known, the percolation algorithm does not make errors with high probability if $K = !(\log^2 n)$. Furthermore, the algorithm percolates to $\frac{n}{K} - o\left(\frac{n}{K}\right)$ vertices in each cluster if $\Phi > \left(1 - \frac{1}{r}\right)\left(\frac{K^r(r-1)!}{n(p+(K-1)q)^r}\right)^{\frac{1}{r-1}}$ and $K = o(\sqrt{n})$.

*Proof*. Let $k \in [K]$. Consider a vertex pair$(i',j') \in C_k^1 \times C_k^2$ and j' ≠ i'. Let $X_{ij}(t)$ be the event that the pair (i; j) is matched at time t. We want to bound the probability of this event, $P(X_{ij}(t))$, conditioned on the fact that before time t only correct pairs were matched. Recall that $M_{ij}(t)$ is the score of pair (i; j') at time t as defined in Section 4.2. We have

$$P(X_{ij'}(t)) \leq P(M_{ij'}(t) = 2, M_{ii'}(t) \leq 2, M_{jj'}(t) \leq 2)$$
$$\leq P(M_{ij'}(t) = 2)$$
$$= P(Bin(t, (ps)^2) + Bin((K-1)t, (qs)^2) = 2)$$
$$\leq (t((ps)^2 + (K-1)(qs)^2))^2$$

Here Bin(t; (ps)$^2$) denotes the Binomial random variable with t trials and success probability (ps)$^2$. An important observation here is that, before time step t, the algorithm has already used t vertex pairs from every community. Hence, to union bound the above probability only $\frac{n}{K}$-t  vertices remain to be considered at time t. Now, we union bound the

$$\bigcup_{k,t,i,j'} P(X_{ij'}(t))$$
$$\leq K \sum_{t=1}^{n/K} \left(\frac{n}{K} - t\right)^2 (t((ps)^2 + (K-1)(qs)^2))^2$$
$$\leq C \frac{\log^4 n}{K^2}$$

(10)

for some constant C, where (10) follows from simple algebraic manipulations. This completes the rst part of the proof. The second part follows in a straight-forward manner from Lemma 12 and Theorem 1 in [4].

The above theorem implies that the number of communities needs to scale just faster than log$^2$ n so that our percolation algorithm does not make errors with r = 2. Without the community label information, the threshold is r = 4 as shown in [44], thus, the community label information is helpful in reducing the seeds required to percolate.

**Two Threshold Percolation Algorithm** A drawback of the basic percolation algorithm [44] is that with r 4, the algorithm requires many seeds to percolate. In fact, the number of seeds required to achieve percolation is an increasing function of the threshold value r. As has been noted in the literature [44, 22], the percolation algorithm makes errors in the initial steps when there are only a few seeds to deanonymize the graphs. Hence, to reduce such errors, we propose a two-threshold percolation algorithm which takes as input an imperfect matching and builds upon this information to percolate further. Assuming that the algorithm percolates, we also derive conditions under which the algorithm does not make errors with high probability.

Consider, as before, two correlated SBM graphs $G_1$ and $G_2$, but we do not assume that we have access to the community labels of the users. Suppose that there exists an algorithm, A, which gives us the following output. For each vertex $I \in V(G_1)$, let $j^i \in V(G_2)$ be the vertex matched to i using the algorithm A. Let us call this set of matched pairs F defined by $F \triangleq \{(i; j^i) \in V(G_1) \times (G_2)$ : A matches i and $j^i\}$. Consider two thresholds, $r_c$ and $r_m$, with $r_c$ $r_m$. For the percolation process, starting with a set of seed pairs, $S_0$, the following strategy is used. Pick a seed pair from the seed set. For vertex i, the pair $(i; j^0)$ is considered as matched as soon as its score, $M_{iji}0$, reaches $r_c$. For all other potential pairs $(I, j)$, the matching threshold is $r_m$. More compactly, the vertex pairs which were matched by algorithm A are given a lower threshold than the other potential pairs. If algorithm A made error, we expect the percolation algorithm to correct them. Using a lower threshold for the pair $(i; j^i)$ is helpful in reducing errors in the initial stages of the percolation process if $j^i = I'$. As the correct pairs are more probable to accumulate marks earlier than incorrect pairs [44], the lower threshold also helps in reducing the required number of seeds. Hence, the more correct pairs algorithm A matches, the fewer seeds are required to percolate and the lower the error in the second step. Algorithm 2 describes the pseudo code of the proposed algorithm.

**Asymptotic Analysis** The similarity of graph matching algorithms to bootstrap percolation is noted in the literature [44]. The latter, in general, studies the spread of infection among vertices in graphs where the infected vertices con-tribute to the infection of their neighbors [18]. When the number of infected neighbors of a vertex reaches a certain threshold, it gets infected as well. The graph matching algorithm can be thought of along similar lines. Instead of ver-tices spreading infection, it is the pair of vertices spreading it in the intersection graph $G_\cap \triangleq G_1 \cap G_2$ with vertex set $V(G_1)$ and edge set $E(G_1) \cap \pi_0^{-1}(E(G_2))$. To make the two problems isomorphic, one rst needs to establish that the percolation algorithm does not make errors whenever it percolates. Therefore, we analyze the proposed algorithm and derive conditions under which the algorithm does not make errors with high probability. In particular, we aim to characterize the performance of algorithm A, required to guarantee that no errors are made while percolating.

---

**Algorithm 2** Percolation Algorithm with Two Thresholds

---

Require: $G_1$; $G_2$; $S_0$; F

  $S = S_0$; $P = \Phi$; $M = 0_{n \times n}$

  while $S \setminus P \neq \Phi$ ; and $|S| \neq n$ do

    Pick a seed pair $(u; v) \in S \setminus P$

    Add $(u; v)$ to P

    for all $(i; j) \in V$ $(G_1) \times V$ $(G_2)$ such that $ui \in E(G_1)$ and $vj' \in E(G_2)$ do

      $M_{ij'} = M_{ij'} + 1$

      if $((j_i = j'$ and $M_{ij} = r_c)\ ||\ (j_i \neq j'$ and $M_{ij} = r_m))$ and both $i; j'$ are unmatched then

        $S = S \cup (i; j)$

        Mark $i$ and $j'$ as matched

      end if

    end for

  end while

  Return S as the set of matched vertices

---

We assume that algorithm A is independent of the percolation step and vice versa. The following theorem provides the conditions under which the algorithm does not make errors with high probability, assuming that it percolates.

**Theorem 3.** *Let $r_m \geq 4$ and $r_c$ 1. Consider any $i$ 2 V $(G_1)$ and $j' \in V$ $(G_2)$ such that $j' \neq i$. For any $(i, j) \in V$ $(G_1) \times V$ $(G_2)$, let $p_f = P$ $((i; j) \in F)$.*
*If $p_f = o(n^{r_c}$ $^{-3}(\log n)$ $^{-2r_c})$, then Algorithm 2 does not match wrong pairs with high probability.*

*Proof.* Consider a vertex pair $(i; j') \in V$ $(G_1) \times$ $(G_2)$ and $j' \neq i'$. Let $X_{ij'}(t)$ be the event that the pair $(i, j')$ is matched at time $t$. We want to bound the probability of this event, $P$ $(X_{ij'}(t))$, conditioned on the fact that before time $t$ only correct pairs were matched. De ne a random variable $f_{ij}$ as

$$f_{ij'} = \begin{cases} 1 & \text{if } j' = j'_i \\ 0 & \text{if } j' \neq j'_i \end{cases}$$

Now, we can bound the above probability as

$$
\begin{aligned}
P(X_{ij'}(t)) &\leq \\
&P(f_{ij'} = 1, M_{ij'}(t) = r_c, M_{ii'}(t) \leq r_m, M_{jj'}(t) \leq r_m) \\
&+ P(f_{ij'} = 0, M_{ij'}(t) = r_m, M_{ii'}(t) \leq r_m, M_{jj'}(t) \leq r_m) \\
&= P(E_1) + P(E_2)
\end{aligned}
\tag{11}
$$

where $E_1$ is the event $[f_{ij'} = 1;\ M_{ij'}(t) = r_c;\ M_{ii'}(t) \leq r_m;\ M_{jj'}(t) \leq r_m]$ and $E_2$ is the event $[f_{ij'} = 0;\ M_{ij'}(t) = r_m,\ M_{ii'}(t) \leq r_m,\ M_{jj'}(t) \leq r_m]$.

Till time $t$, let $T_1$, and $T_2$ be the random variables corresponding to the number of seeds with community labels $C(i)$ and $C(j')$ respectively. Consider that $C(i) \neq C(j')$. The other case, when $C(i) = C(j')$, follows similarly by substituting $T_2 = 0$ and replacing $pq$ by $p^2$ in the equations below. We have

$$P(E_2) \overset{(a)}{\leq} P[M_{ij'}(t) = r_m]$$

$$\overset{(b)}{=} \sum_{t_1=0}^{t} \sum_{t_2=t-t_1}^{t} \big[ P(M_{ij'}(t) = r_m | T_1 = t_1, T_2 = t_2)$$

$$P(T_1 = t_1, T_2 = t_2) \big]$$

$$= \sum_{t_1=0}^{t} \sum_{t_2=t-t_1}^{t} \big[ P(Bin(t_1 + t_2, pqs^2) +$$

$$Bin(t - t_1 - t_2, (qs)^2) = r_m) P(T_1 = t_1, T_2 = t_2) \big]$$

$$\overset{(c)}{\leq} \sum_{t_1=0}^{t} \sum_{t_2=t-t_1}^{t} \big[ P(Bin(t, pqs^2) = r_m)$$

$$P(T_1 = t_1, T_2 = t_2) \big]$$

$$\leq (tpqs^2)^{r_m} \overset{(d)}{\leq} (npqs^2)^{r_m}$$

$$= (ab)^{r_m} s^{2r_m} n^{-r_m} (\log(n))^{2r_m}$$

$$\vdots$$

(12)

where (a) is true because the probability of the intersection of multiple events is smaller than the probability of individual events, (b) follows from the law of total probability, (c) follows from $q < p$ and because $n(qs)^2 \to 0$, and (d) follows from $t \leq n$. If $r_m \geq 4$, then

$$\bigcup_{i,j',t} P(E_2) \leq (ab)^{r_m} s^{2r_m} n^{3-r_m} (\log(n))^{2r_m} \to 0$$

(13)

For event $E_1$, we have

$$P(E_1) \leq P[f_{ij'} = 1, M_{ij'}(t) = r_c]$$

$$= P[f_{ij'} = 1] \times P(M_{ij'}(t) = r_c)$$

$$\leq p_f (tpqs^2)^{r_c}$$

Using union bound and the condition in the theorem, we have

$$\bigcup_{i,j',t} P(E_1) \leq p_f (ab)^{r_c} s^{2r_c} n^{3-r_c} (\log(n))^{2r_c} \to 0$$

(14)

When the community label information is available, the naive algorithm pro-posed earlier can be used as a preprocessing algorithm before percolation. In the next section, we show experimentally that Algorithm 2, with $r_c = 1$ and $r_m = 2$, performs much better than similar graph matching algorithms. This in turn shows that the community labels serve as excellent side information, helping both in reducing the number of seeds needed to percolate and achieving lower error rates.

# 5 Experimental Evaluation

In this section, we evaluate the proposed algorithms on synthetic as well as real world datasets. The main goal of this section is to show that information about community labels, if used properly, enhances the deanonymization results especially when the correlation between the networks is low.

In the experiments below, we compare the following algorithms:

- **A1**: percolation with $r_c = r_m = 1$
- **A2**: percolation with $r_c = 1$, $r_m = 2$ (our proposed algorithm)
- **A3**: percolation with $r_c = r_m = 2$
- **A4**: ExpandWhenStuck Algorithm [22]

Algorithms **A1** and **A3** correspond to percolation graph matching with uni-form thresholds [44], but match vertices belonging to the same community only. Algorithm **A4** is the ExpandWhenStuck algorithm proposed in [22]. To the best of our knowledge, **A4** achieves the best time-accuracy trade-off for matching large graphs. Algorithm **A2** is the proposed algorithm which uses two different thresholds for matching the graphs. We use the naive algorithm as A. We eval-uate the algorithm for $r_c = 1$ and $r_m = 2$, which achieves excellent results with minimal number of seeds.

We compare the performance of the algorithms in terms of percolation strength and error rate with varying numbers of seeds ($\Phi$) and values of sampling parameter (s). For all the experiments, the seeds are chosen uniformly at ran-dom. To measure the percolation strength, we compare the fraction, f, of vertices matched by the algorithms, i.e., $f = \frac{\# \, vertices \, matched}{n}$. The error rate, e, is measured by the fraction of vertices matched incorrectly, i.e., e $= \frac{\# \, incorrect \, matches}{\# \, vertices \, matched}$.

## 5.1 Description of Datasets

We evaluate the proposed algorithm on synthetic as well as real world datasets. For the synthetic networks, the underlying graph, G, is generated using the Stochastic Block Model with parameters n = 10000, and b = 2 (unless stated otherwise). The real world datasets are the collaboration networks of authors who submitted papers to Astro Physics (As-Physics) and Condensed Matter Physics (CondMat) categories on arXiv [26]. In these networks, two authors are connected if they co-authored a paper. We consider only the largest connected component in these graphs. The As-Physics datasets consists of 17903 users and 196972 edges, the average degree being 22:01. For the CondMat dataset, these numbers are 21363, 91286, and 8:55 respectively. We use Louvain community detection algorithm implemented in Pajek on the underlying graph to detect the community structure [6]. We detect 36 and 55 communities in the As-Physics and CondMat datasets, respectively. To generate correlated networks, $G_1$ and $G_2$, we sample the edges in the underlying graphs with probability s (values of s are described in subsequent subsections). As we aim to demonstrate the importance of community labels in graph matching, we assume that the attacker knows the community labels of the vertices.

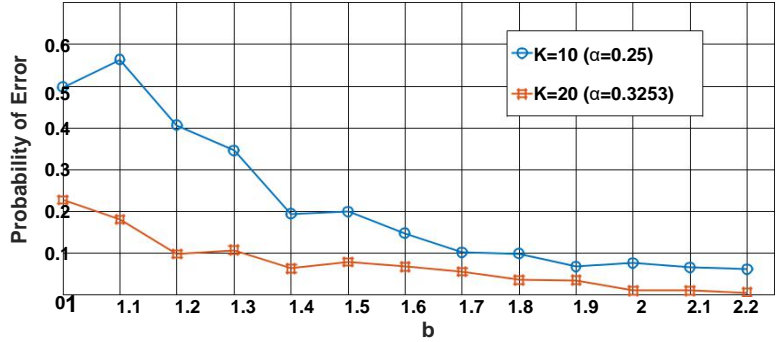

Fig. 2: Probability of error versus parameter b for SBM graphs with s = 0.7.

### 5.2 Probability of Error

Although the results derived in this paper hold asymptotically, it is important to assess their validity on nite graphs. Let $p_{err}$ denote the probability that the proposed percolation algorithm A2 makes an error. To evaluate this probability, we generate 200 random instances of the problem and match graphs $G_1$ and $G_2$ using the proposed algorithm. Let $n_{err}$ denote the number of instances when the proposed algorithm makes an error. Thus, $p_{err} = \frac{n_{err}}{200}$. Figure 2 shows the obtained results as a function of parameter b for two values of K and s = 0.7. For small values of b, $p_{err}$ is high and it decreases as b increases. For large values of b and K = 10, the algorithm attains $p_{err}$ of around 0.08 while for K = 20, $p_{err}$ is almost zero. This shows that side information in the form of community labels is extremely helpful in matching the graphs.

### 5.3 Effect of Number of Communities

We first measure the of number of communities on the proposed naive (Algorithm 1) and percolation (A2) algorithms. Figure 3 shows the error rate achieved by the naive algorithm for SBM graphs with different values of the sam-pling parameter. It can be noted that as the number of communities increases, the error rate decreases. For larger values of s, even a very small number of communities is enough to deanonymize a large percentage of users.

In Figure 4, we depict the performance of algorithm A2 for s = 0.5. For K = 10, the algorithm requires around 245 seeds to percolate as virtually all the correct pairs of vertices get a higher threshold using the naive algorithm in the rst step (see Figure 3). However, for K = 20, the number of seeds required to percolate decreases drastically to 35 even though the fraction of correct pairs assigned a lower threshold is less than 10%. As the number of communities increases further, a greater fraction of correct pairs get lower threshold and hence percolation is immediate for larger values of K. This shows that the performance of our algorithm improves with the number of communities as expected. With only 20 communities, the algorithm achieves less than 0.4% error rate although the correlation between the networks is very weak.

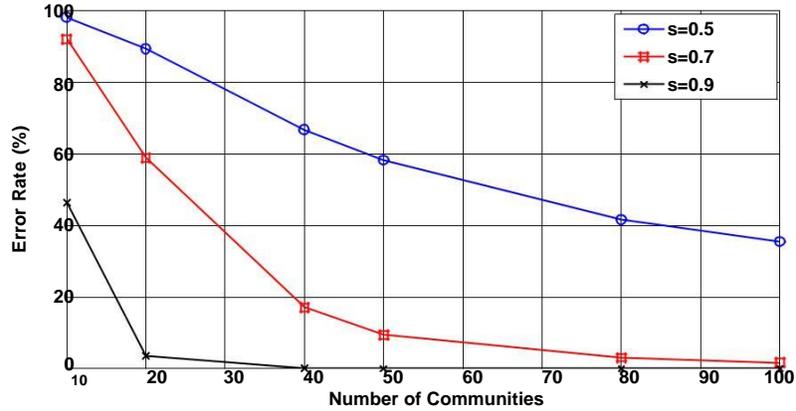

Fig. 3: Effect of number of communities on the error rate of the naive algorithm for SBM graphs with parameters n = 10000, b = 2.

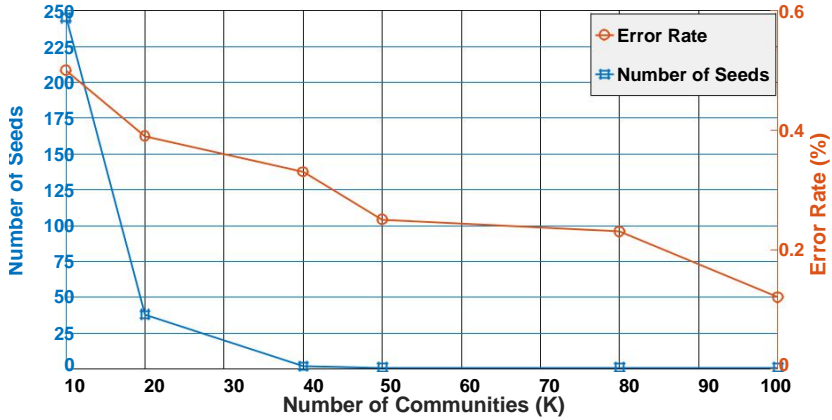

Fig. 4: Effect of number of communities on the performance of the proposed percolation algorithm for SBM graphs with parameters n = 10000, b = 2, s = 0.5.

### 5.4 Effect of Seeds

First, we compare the variation in the performance of algorithms A2 and A4 by varying the initial seed set $S_0$. We use 100 random seed sets of size 5 to obtain the results. Figure 5 shows the comparison of error rate distributions for three datasets. It can be seen that our algorithm is insensitive whereas algorithm A4 is highly sensitive to the quality of seeds. As the median is robust to outliers, we report the median of error rates over 100 random seed sets for algorithm A4 in the experiments.

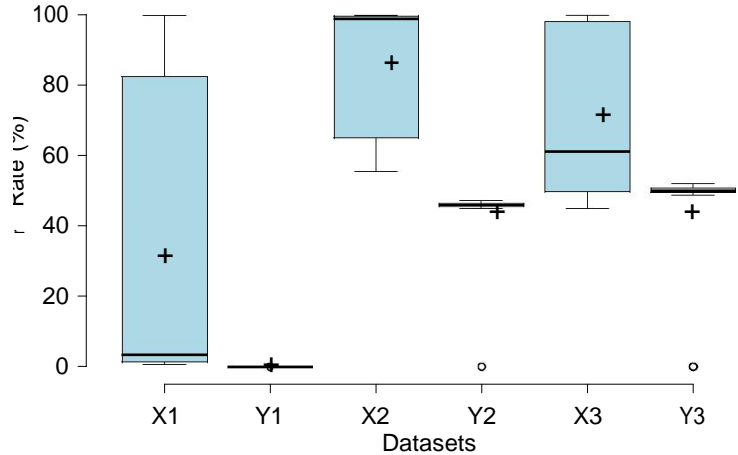

Fig. 5: Box plot comparing distributions of error rate for algorithms A2 and A4 for 100 random seed sets of size 5. X and Y represent algorithms A4 and A2 respectively. Suffixes 1,2, and 3 represent datasets, 1: SBM (s = 0.8; K = 20), 2: CondMat (s = 0.7), 3: As-Physics (s = 0.7). Box represents the Inter Quartile Range (IQR), horizontal line in the box denotes the median, +, denote mean and outliers respectively, horizontal lines above and below the box denote maximum and minimum respectively.

Figure 6 shows the deanonymization results for the synthetic networks with s = 0:8 and K = 20 for varying number of seeds. Figure 6(a) depicts the perco-lation behavior of the algorithms. Algorithm A3 is dropped because it required more seeds to percolate than the other three algorithms. It can be seen that all three algorithms percolate instantaneously, but the proposed algorithm A2 maps almost the complete network. Note that algorithm A4 achieves the same per-centage of correctly matched vertices with 10 seeds. The error rate comparison is shown in Figure 6(b). The proposed algorithm outperforms both algorithms A1 and A4 and achieves near zero error rate with only 1 seed. It is interesting to note that algorithm A4 is able to perform well with around 5 seeds and is worse than A1 for fewer seeds.

Figure 7 shows the deanonymization results for the As-Physics dataset with s = 0.9. The fractions of users in the intersection with degree greater than or equal to 1 and 2 are approximately 98% and 90% respectively. The community

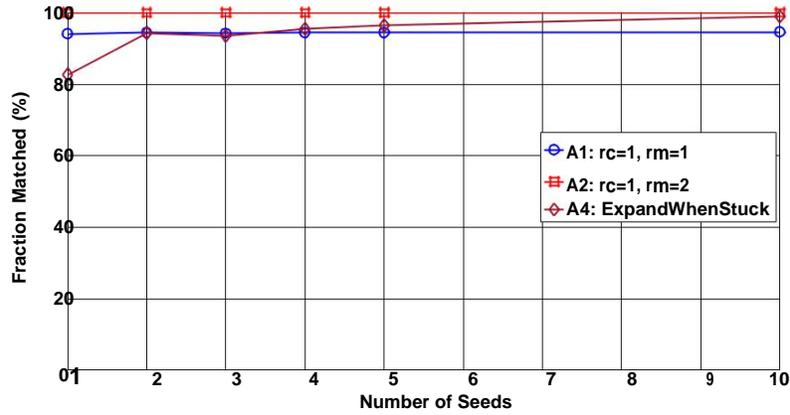

*(a) Fraction of vertices matched vs Number of seeds*

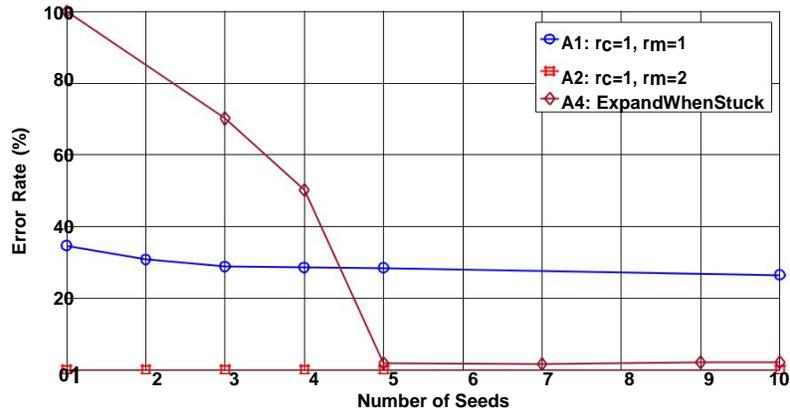

*(b) Error Rate vs Number of Seeds*

Fig. 6: Deanonymization results for SBM dataset with parameters n = 10000, b = 2, s = 0.8, and K = 20.

detection algorithm partitions the network into 36 disjoint communities. It can be noted that using constant threshold $r_c = r_m = 1$, algorithm A1 maps around 83% of the users but results in very high error rate. Algorithm A3 needs around 9 seeds to percolate to 72% of the users. The error rate in this case is more than 40%. The proposed algorithm performs much better than the other two. It maps more than 82% users with error rate around 24%. Also note that our algorithm percolates with only 2 seeds. Although algorithm A4 maps more users, our algorithm achieves lower error rate.

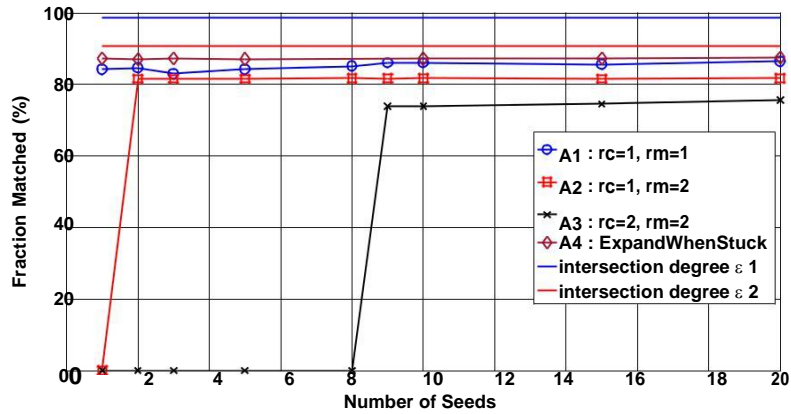

*(a) Fraction of vertices matched vs Number of seeds*

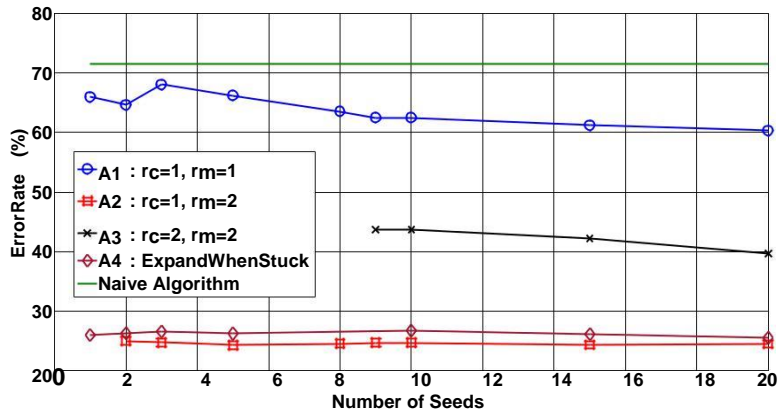

*(b) Error Rate vs Number of Seeds*

Fig. 7: Deanonymization results for As-Physics dataset with s = 0.9.

Figures 8 shows the deanonymization results for the CondMat dataset, which is much sparser than the previous dataset, for s = 0.7. The fractions of users with degrees at least 1 and 2 in the intersection graph are around 96% and 70% respectively. Algorithm A1 resulted in more than 80% error rate and algorithm A3 needed more than 20 seeds to percolate hence the results of these algorithms are not shown. The other two algorithms percolate with only 2 seeds, as can be seen in Figure 8(a), with algorithm A4 mapping slightly higher number of users than the proposed algorithm. However, as depicted in Figure 8(b), the former algorithm makes more errors than our algorithm. Figure 8(b) also shows the Inter

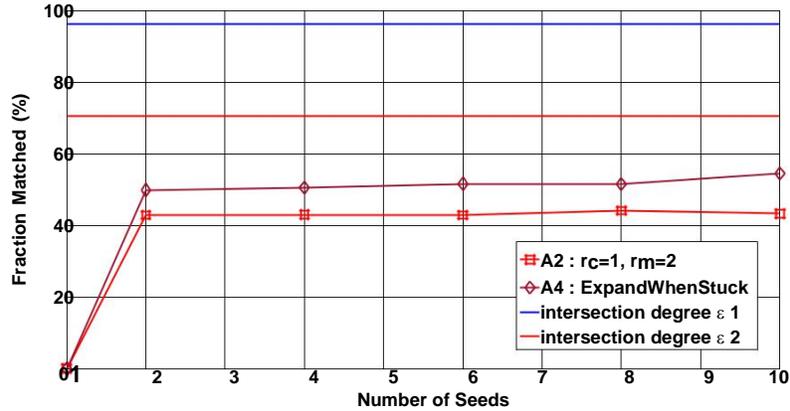

*(a) Fraction of vertices matched vs Number of seeds*

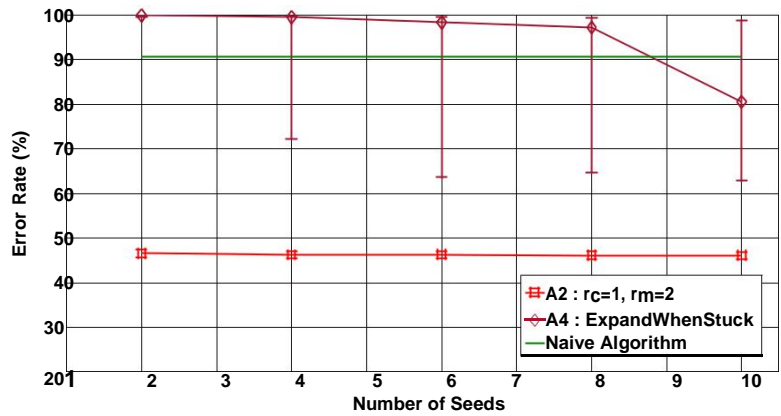

*(b) Error Rate vs Number of Seeds*

Fig. 8: Deanonymization results for CondMat dataset with s = 0.7.

Quartile Range (IQR) for algorithm A4. Note that even the 25[th] percentile is well above the error rate achieved by the proposed algorithm, even with 10 seeds.

### 5.5 Effect of Sampling Parameter

In this subsection, we compare the performance of algorithms by varying the levels of correlation. In the rst experiment, we compare the percolation threshold (minimum number of seeds required to percolate) of algorithms A2 and A3 for the synthetic dataset with K = 10, 20. As seen in Figure 9, the proposed algorithm requires far fewer seeds to percolate than the percolation algorithm with uniform threshold. In particular, for s = 0.5, assigning a lower threshold to less than 10% correct pairs (see Figure 3) changes the percolation threshold from 285 to 35 for K = 20. At higher values of s, algorithm A2 requires only 1 seed to percolate. Also note that for K = 20, the algorithm requires fewer seeds than for K = 10. The reason for this is the fact that when K = 20, more correct pairs get a lower threshold for matching. This highlights the importance of community labels as side information, especially when number of communities is bigger.

In the next set of experiments, we compare the overall performance of the algorithms A1-A4 on synthetic and real datasets. The number of seeds is xed at 10 for all the experiments. To measure the performance, we use the $F_1$-score defined as follows:

$$F_1 = 2 \frac{P \, R}{P + R} \qquad (15)$$

where P = 1    e, R = $\frac{\#correct\ matches}{n_{int}}$ are the precision and recall respectively.

Here $n_{int}$ is the number of vertices in the giant component of the intersection graph $G_n$. The $F_1$ score combines both the percolation strength and error rates achieved by the algorithms into a single index. An $F_1$ score closer to 1 means better performance of an algorithm in terms of both percolation strength and error rate.

Figure 10 shows the $F_1$ scores for the synthetic dataset with K = 20. Algorithm A3 is not shown as it did not percolate for any value of s. Note that our algorithm outperforms algorithm A1 for most values of s; at s = 0.5 our algorithm did not percolate. Compared to algorithm A4, our algorithm achieves better scores when the datasets are weakly correlated (s ≤ 0.7). Even when s = 0.8, our algorithm is slightly better than algorithm A4.

Figure 11 shows the $F_1$ scores for the As-Physics dataset. As before, for low values of s, the proposed algorithm outperforms the other three algorithms. Algorithm A1 percolates, but makes many errors and hence achieves low scores even at higher values of s. At s = 0.9, the proposed algorithm obtains a score which is better by a value of 0.1 than that of algorithm A4. Note that the IQR at s = 0.7 for A4 is large, which shows the sensitivity of the algorithm to the quality of seeds.

Figure 12 shows the $F_1$ scores for the CondMat dataset. At high values of s, algorithm A4 achieves slightly better scores than the proposed algorithm. However, at lower values, the proposed algorithm outperforms other algorithms. It is worth noting that algorithm A4 results in almost 100% error rate for s 0.6, although it percolates.

# 6 Conclusions

In this paper, we investigate the importance of side information in the graph matching problem. We show that the information about the community structure can be used to match two correlated SBM graphs perfectly with high prob-ability, without using seeds. When the seeds are available, we also show that the community structure helps in reducing the matching threshold, in the percolation graph matching algorithm, from r = 4 to r = 2. When an attacker has access to an imperfect matching, we propose a novel percolation algorithm with two thresholds, to match two correlated graphs. We prove the conditions under which the algorithm does not make errors whenever it percolates. Experimental results suggest that our algorithm performs extremely well on the synthetic and real world networks and outperforms other percolation algorithms when the cor-relation between the networks is low. We also show that our algorithm percolates with only a few seeds, is robust to the quality of seeds and works well even when the community labels are known only for a fraction of users.

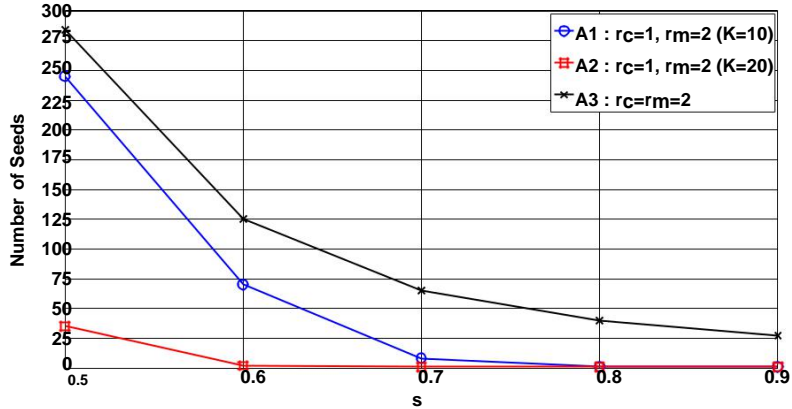

Fig. 9: Percolation threshold for algorithms A2 and A3.

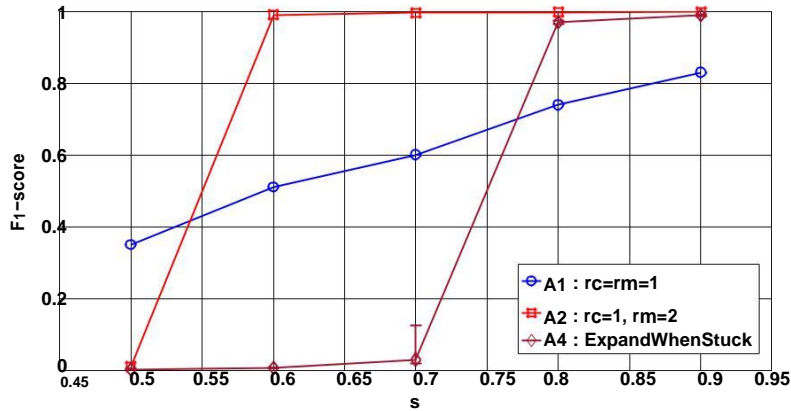

Fig. 10: F$_1$-score comparison for different levels of correlation for the synthetic dataset.

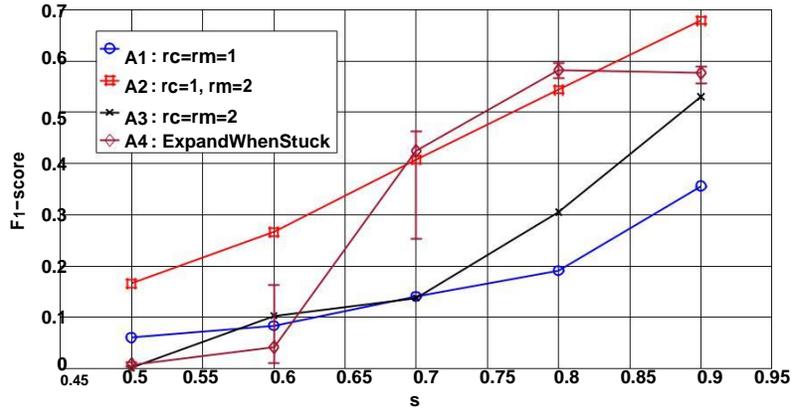

Fig. 11: F$_1$-score comparison for different levels of correlation for the As-Physics dataset.

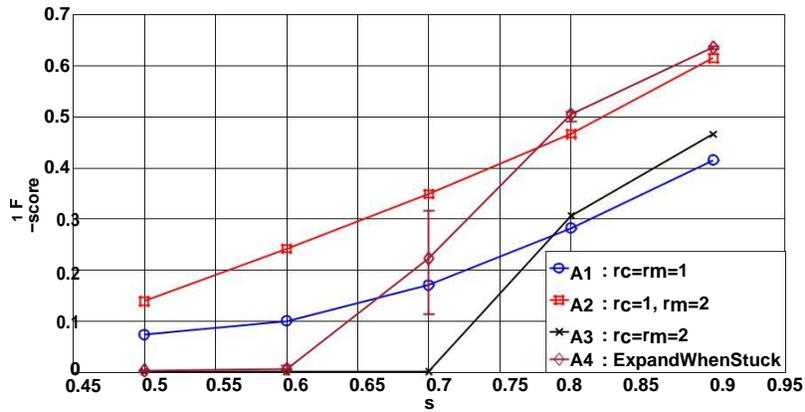

Fig. 12: F$_1$-score comparison for different levels of correlation for the CondMat dataset.

There are two main practical implications of our work. First, it shows that side information in the form of community labels has the capability to assist an attacker in network deanonymization, especially in the low correlation regime. Second, in situations where the correlation is strong, information about community labels helps in reducing the number of seeds needed to percolate and mitigates the Effect of seed quality on the deanonymization results.

There are various future directions to explore as a result of our findings. For example, it is pertinent to study other types of side information which could help an attacker to match correlated networks. Such studies could help in designing mechanisms to mitigate the Effect of such side information in network deanonymization. Studying similar two-threshold percolation algorithms for other random graph models like preferential attachment [5], and con guration models [33], can provide more insights into the problem of matching real world graphs. A more realistic modi cation of this problem involves considering overlapping communities. We expect our results to motivate research in these directions in order to understand the graph matching problem more concretely.